\documentclass{aip-cp} 
\pdfoutput=1
\usepackage{graphicx} 
\usepackage{amssymb} 
\usepackage{amstext}
\usepackage{amsfonts} 
\usepackage{slashed} 
\usepackage[dvipsnames]{xcolor} 

\newcommand{\nopieft}{\mbox{$\slashed{\pi}$EFT}}

\newcommand{\be}{\begin{equation}} 
\newcommand{\ee}{\end{equation}}

\newcommand{\rvec}{{\vec{r}}}

\newcommand{\half}{\frac{1}{2}}

\begin{document}

\title{B$_\Lambda$($^5_\Lambda$He) from short range effective field theory}
\author[1]{L.~Contessi\corref{cor1}} 
\author[1]{N.~Barnea} 
\author[1]{A.~Gal}
\affil[1]{Racah Institute of Physics, The Hebrew University, Jerusalem 91904, 
Israel} 
\corresp[cor1]{Corresponding author: contessi.lorenzo@mail.huji.ac.il}

\maketitle

\begin{abstract}
We present an effective field theory (EFT) at leading order to describe light 
single-$\Lambda$ hypernuclei. Owing to the weak $\Lambda$ binding and to the 
$\Lambda N$ short interaction range, meson exchange forces are approximated by 
contact interactions within a pionless EFT (\nopieft) where the only degrees 
of freedom are baryons. At leading order, the $\Lambda$-nuclear interaction 
contains two 2-body (singlet and triplet) and three 3-body interaction terms, 
a total of 5 terms associated with 5 coupling strengths or low energy 
constants (LECs). We adopt the 2-body LECs from hyperon-nucleon scattering 
data and interaction models that constrain the $\Lambda N$ scattering lengths, 
while the 3-body LECs are adjusted using both 3-body and 4-body hypernuclear 
binding energies. To calculate the binding energies for A-body systems with 
A$>$2, we expand the wavefunctions using a correlated Gaussian basis. The 
stochastic variational method is employed to select the non-linear parameters. 
The resulting \nopieft~is then applied to calculate the $\Lambda$ separation 
energy in $_\Lambda^5$He, where the adjusted 3-body interactions largely 
resolve the known overbinding problem of $_\Lambda^5$He. 
\end{abstract}

\section{INTRODUCTION} 

The inclusion of a $\Lambda$ particle in nuclei is the first natural step in 
extending the periodic table into the strangeness sector. While other hyperons 
such as $\Sigma$ and $\Xi$ might be considered in a theoretical framework, the 
available hypernuclear data consist almost exclusively of single-$\Lambda$ 
hypernuclei which present some fascinating questions, and complications, 
for theory to resolve. In the present work, we aim to address two of several 
unsolved problems in $\Lambda$ hypernuclei: (i) the difficulties in 
establishing precise $\Lambda N$ scattering parameters from experimental 
results, and (ii) the so called \textit{overbinding problem} of $_\Lambda^5$He 
in modern $\Lambda$-nuclear interactions.

\begin{table}[!ht] 
\caption{Input scattering lengths (in fm) used to adjust \nopieft~LECs. 
Except for the Nijmegen models (based on model NSC97f~\cite{NSC97}) the 
listed $\Lambda N$ scattering lengths are identified with tabulated 
$\Lambda p$ scatering lengths.} 
\label{tab:LN} 
\begin{tabular}{cccccc} 
\hline 
$\Lambda N$ Fit/Model & Ref. & $a_s(NN)$ & $a_s(\Lambda N)$ & $a_t(\Lambda N)$ 
& ${\bar a}_{\Lambda N}$  \\
\hline 
Alexander[A] & \cite{Alex68} & $-$23.72 & $-$1.8 & $-$1.6 & $-$1.65  \\
Alexander[B] & \cite{Alex68} & $-$18.63 & $-$1.8 & $-$1.6 & $-$1.65  \\
Nijmegen[A]       & \cite{NSC97}  &  $-$23.72 & $-$2.60 & $-$1.71 & $-$1.93  \\ 
Nijmegen[B]       & \cite{NSC97}  & $-$18.63 & $-$2.60 & $-$1.71 & $-$1.93  \\ 
$\chi$EFT(LO) & \cite{Polinder06} & $-$18.63 & $-$1.91 & $-$1.23 & $-$1.40  \\ 
$\chi$EFT(NLO) & \cite{Haiden13} & $-$18.63 & $-$2.91 & $-$1.54 & $-$1.88  \\ 
\hline 
\end{tabular} 
\end{table}

The experimental difficulties in measuring $\Lambda N$ scattering parameters 
stem from the unavailability of hyperon beams or hyperon targets, thereby 
limiting all measurements to the use of secondary interactions. Most of 
the data available in this 2-body sector, which is of great importance 
for theory, consist of spin-independent $\Lambda p$ scattering total 
cross sections at insufficiently low energies \cite{Alex68,sechi68}. 
Table \ref{tab:LN} demonstrates the uncertainty exhibited by adopting 
$\Lambda N$ scattering lengths directly from fits to $\Lambda p$ scattering 
cross sections~\cite{Alex68} or from several leading hyperon-nucleon (YN) 
interaction models, with spin-singlet $a_s(\Lambda N)$ values varying between 
$-$1.91 and $-$2.91 fm and spin-triplet $a_t(\Lambda N)$ values varying 
between $-$1.23 and $-$1.71 fm. Interestingly, the uncertainty in the 
listed values of the spin-independent scattering length combination 
${\bar a}_{\Lambda N}=(3/4)a_t +(1/4)a_s$ is somewhat smaller. 
Hypernuclear data too are not as abundant as nuclear data are. 
The known $\Lambda$ hypernuclear binding energies are limited to a few 
dozens of systems, for many of which the deduced information is further 
limited to ground states. In spite of several recent experiments on light 
hypernuclei~\cite{MAMI16A,MAMI16B,E13} our knowledge in this sector remains 
incomplete. Altogether one is far from having the precision and extension 
of experimental data that is found in standard nuclear physics. 

\begin{table}[!ht] 
\caption{Ground-state $\Lambda$ separation energies $B_{\Lambda}$ and 
excitation energies $E_x$ (in MeV) from several few-body calculations 
of $s$-shell $\Lambda$ hypernuclei, see text. Charge symmetry breaking 
is included in the $_\Lambda^4$H results from Ref.~\cite{GG16A,GG16B}.}  
\begin{tabular}{ccccc} 
\hline 
 & $B_{\Lambda}(_\Lambda^3$H) & $B_{\Lambda}(_\Lambda^4$H$_{\rm g.s.}$) 
& $E_x(_\Lambda^4$H$_{\rm exc.}$) & $B_{\Lambda}(_\Lambda^5$He)  \\ 
Exp. & 0.13(5)~\cite{Davis05} & 2.16(8)~\cite{MAMI16A,MAMI16B} & 
1.09(2)~\cite{E13} & 3.12(2)~\cite{Davis05}  \\
\hline 
Dalitz et al.~\cite{DHT72} & 0.10 & 2.24 & 0.36 & $\geq$5.16  \\ 
AFDMC (I) & -- & 1.97(11)~\cite{LPG14} & -- & 5.1(1)~\cite{LGP13}  \\ 
AFDMC (II) & $-$1.22(15)~\cite{LPG14} & 1.07(8)~\cite{LPG14} & -- & 
3.22(14)~\cite{LPG14}  \\ 
AFDMC (III) & 0.23(9)~\cite{LPL18} & 1.95(9)~\cite{LPL18} & -- & 
2.75(5)~\cite{LPL18}  \\ 
$\chi$EFT(LO$_{600}$) & 0.11(1)~\cite{Wirth14} & 2.31(3)~\cite{GG16A,GG16B} & 
0.95(15)~\cite{GG16A,GG16B} & 5.82(2)~\cite{WR18}  \\  
$\chi$EFT(LO$_{700}$) & -- & 2.13(3)~\cite{GG16A,GG16B} & 
1.39(15)~\cite{GG16A,GG16B} & 4.43(2)~\cite{WR18}  \\ 
\hline 
\end{tabular} 
\label{tab:over} 
\end{table} 

It is not surprising, given this background, that many interaction models 
have been formulated to describe hypernuclei. Most of these models describe 
well the few-body (A$\leq$4) sector but overbind heavier hypernuclei starting 
from $^5_\Lambda$He which is overbound by 1-2 MeV and for which precise 
ab initio calculations are still possible. In Table \ref{tab:over} 
several of the most commonly used interaction models are listed together with 
their resulting ground-state $\Lambda$ separation energies $B_{\Lambda}$ and 
excitation energies $E_x$ for the relevant 3-, 4-, and 5-body hypernuclei. 
Already in 1972 Dalitz et al. \cite{DHT72} realized that $^5_\Lambda$He was 
overbound by using a phenomenological $\Lambda N$ + $\Lambda NN$ model, and 
this overbinding problem has persisted in modern $\chi$EFT interaction models 
\cite{GG16A,GG16B,WR18} at leading order (LO) and for a wide range of cut-off 
values (calculations for NLO in $\chi$EFT have not been reported yet). The 
only published calculation which seems not to be plagued by the overbinding 
problem is the one using the AFDMC (II) interaction model \cite{LPG14} which 
is based on a Bodmer-type $\Lambda N$ interaction \cite{BUC84} and on refitted 
Urbana-like $\Lambda NN$ interactions. However, this interaction model 
underbinds the 3- and 4-body systems, thus shifting the $^5_\Lambda$He 
overbinding problem to an underbinding problem for the lighter hypernuclei. 
Recently these authors presented a version of the same interaction, AFDMC 
(III) \cite{LPL18}, in which the $\Lambda$ separation energies in the 3- and 
4-body systems are well reproduced and $^5_\Lambda$He is even underbound, 
implying that $^5_\Lambda$He still requires to be fully understood. These 
calculations suggest that a 2-body $\Lambda N$ interaction is indeed 
insufficient to reproduce hypernuclear ground-state separation energies, 
but also that the mere introduction of 3-body $\Lambda NN$ interactions 
does not guarantee that $B_{\Lambda}(_\Lambda^5$He) is reproduced. 

The aim of this report is to show how pionless EFT may be used within 
error-controlled ab initio few-body calculations to come close to a good 
reproduction of $B_\Lambda$($^5_\Lambda$He) without incurring substantial 
overbinding. To this end we review and expand on our recent application of 
\nopieft~to single $\Lambda$ hypernuclei~\cite{mypaper}. In the following, 
$B_\Lambda$($^5_\Lambda$He) is calculated for all of the input two-body 
scattering parameters shown in Table \ref{tab:LN}, thereby linking binding 
energy calculations of $s$-shell hypernuclei directly with $\Lambda N$ 
scattering data and $YN$ model predictions.

\section{PIONLESS EFT}

Effective field theories rely on the relevant symmetries of the underlying 
interactions for the phenomena of interest: Quantum Chromo Dynamics for 
nuclear and hypernuclear physics. In the present problem, the applicable 
degrees of freedom are baryons, which along with the low values of the 
exchanged momentum $Q$ involved justifies the use of a non-relativistic 
approach. Moreover, since in light (A$\leq$5) nuclear and hypernuclear 
systems $Q$ is small, explicit meson exchange plays an insignificant 
role while contact (or pionless) potentials become more appropriate. This 
is especially true for $\Lambda$ hyperons which are weakly bound in light 
hypernuclei, but also for standard nuclei with A$\leq$4 where the contact 
approach proved to give exceptionally good results despite the less clear 
separation of scales \cite{BCG15}. Here we use a regular \nopieft~approach 
for the nuclear interaction as described by van Kolck in \cite{Kol99} 
with two 2-body and one 3-body free parameter and, further, develop 
a \nopieft~approach for hypernuclear systems by adding two 2-body 
$\Lambda N$ and three 3-body $\Lambda NN$ interaction terms. 

In both cases, nuclei and hypernuclei, \nopieft~is applied within the same 
procedure: the interaction at LO assigns one 2-body or a 3-body contact 
term for each possible 2- and 3-body $s$-wave ($L$=0) state. The $NNN$ 
and $\Lambda NN$ contact terms may be viewed as arising dominantly from 
$NN\leftrightarrow\Delta N$ and $\Lambda N\leftrightarrow\Sigma N$ coupled 
channel interactions, respectively, promoted from subleading order to LO.  
Momentum dependent operators, spin-orbit and tensor force, which also appear 
at subleading order in this approach, are not included in the calculation, 
as well as the Coulomb interaction. The free parameters of the theory are the 
LECs which are directly related to the structure of the possible few-body 
states, and are included in the theory as strengths of zero-range contact 
interactions. However, because a zero-range interaction is too singular to 
be used without introducing a regularization/renormalization scheme, it is 
customary to introduce a Gaussian regulator specified by by its momentum 
cut-off $\lambda$ (see e.g. \cite{Bazak16}): 
\begin{equation} 
\delta_\lambda(\rvec)=\left(\frac{\lambda}{2\sqrt{\pi}}\right)^3\,
\exp \left(-{\frac{\lambda^2}{4}}\rvec^{~2}\right), 
\label{eq:gaussian} 
\end{equation} 
thereby making the LECs cut-off dependent. Choosing other local or nonlocal 
regulators affects mostly the fitted LECs. The resulting LO two-baryon 
interaction reads then: 
\begin{equation}  
V_{2B} = \sum_{IS}\,C_{\lambda}^{IS} \sum_{i<j} {\cal P}_{IS}(ij)
          \delta_\lambda(\rvec_{ij}), 
\label{eq:V2} 
\end{equation} 
where ${\cal P}_{IS}$ are projection operators on $NN,\Lambda N$ pairs with 
isospin $I$ and spin $S$ and the coefficients $C_{\lambda}^{IS}$ are the 
respective LECs. The LO three-body interaction consists of a single $NNN$ 
term associated with the $IS$=$\frac{1}{2}\frac{1}{2}$ channel, and three 
$\Lambda NN$ terms associated with the $IS$=$0\frac{1}{2},1\frac{1}{2},
0\frac{3}{2}$ $s$-wave configurations, with explicit forms given by 
\begin{equation} 
V_{NNN} = D_{\lambda}^{\half\half}\sum_{i<j<k}{\cal Q}_{\half\half}(ijk)
\left(\sum_{\rm cyc.}\delta_\lambda(\rvec_{ik})\, \delta_\lambda(\rvec_{jk})
\right),
\label{eq:V3NNN}
\end{equation} 
\begin{equation} 
V_{\Lambda NN} = \sum_{IS}\,D_{\lambda}^{IS} \sum_{i<j}\,
{\cal Q}_{IS}(ij\Lambda)\,\delta_\lambda(\rvec_{i\Lambda})\,
\delta_\lambda(\rvec_{j\Lambda}),
\label{eq:V3NNL} 
\end{equation} 
where the first sum in Eq.~(\ref{eq:V3NNN}) runs over all $NNN$ triplets, 
the second sum in Eq.~(\ref{eq:V3NNL}) runs over all $NN$ pairs, 
${\cal Q}_{IS}$ are projection operators on baryon triplets with isospin 
$I$ and spin $S$, and $D_{\lambda}^{IS}$ denote the corresponding LECs. 

The two nuclear 2-body LECs were fitted here to the deuteron binding energy 
(2.22 MeV) and two different spin-singlet $pn$ scattering lengths in the cases 
Alexander[A] - Alexander[B] and Nijmegen[A] - Nijmegen[B], introduced to test 
the resilience of the hypernuclear theory against small changes in the nuclear 
input. $\Lambda N$ parameters are fitted to the 2-body scattering lengths 
shown in Table \ref{tab:LN}. All the other parameterizations have different 
$\Lambda N$ scattering lengths as listed in Table \ref{tab:LN}. Two of the 
3-body coefficients were fitted to reproduce $B(^3$H) and $B(^3_\Lambda$H), 
but to determine the two remaining $\Lambda NN$ LECs it is not possible 
to use any other 3-body system because of the lack of experimental data. 
Therefore, $B_{\Lambda}(^4_\Lambda$H$_{\rm g.s.}$) and $B_{\Lambda}
(^4_\Lambda$H$_{\rm exc.}$) for the two known levels of the $A=4$ isodoublet 
hypernuclei with spins $S=0,1$, respectively, have been used. $B(^4$He) and 
$B_{\Lambda}(^5_\Lambda$He) emerge then as predictions of the theory. 

In hypernuclei, a one-pion exchange in the $\Lambda N$ interaction is 
forbidden by isospin conservation, making two-pion exchange the longest 
range meson exchange possible and thereby defining an energy breaking scale 
of $2m_{\pi}$, higher than the scale $m_{\pi}$ in the nuclear case. Therefore, 
we expect a truncation error of order $(Q/2m_{\pi})^2\sim$ 9\% for the 
theoretical prediction of $\Lambda$ separation energies in single $\Lambda$ 
hypernuclei, where the momentum scale $Q$ is provided in light $\Lambda$ 
hypernuclei by $Q\sim p_{\Lambda}\approx\sqrt{2M_{\Lambda}B_{\Lambda}^{
\rm exp.}({_{\Lambda}^5}{\rm He})}=83$~MeV/c.

\section{RESULTS}

\begin{figure}[hpt!] 
\includegraphics[width=0.5\textwidth]{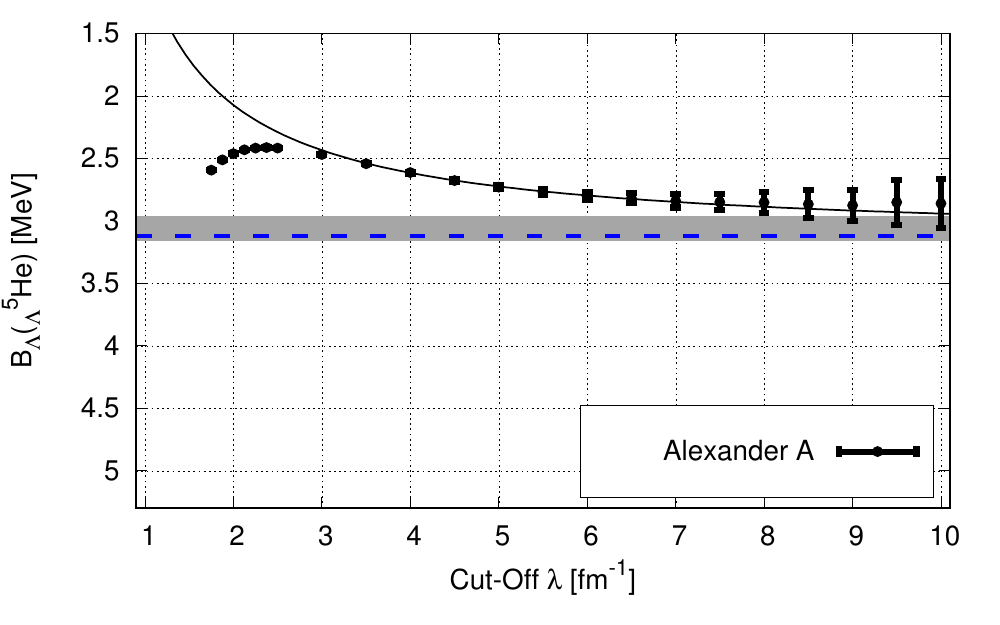} 
\includegraphics[width=0.5\textwidth]{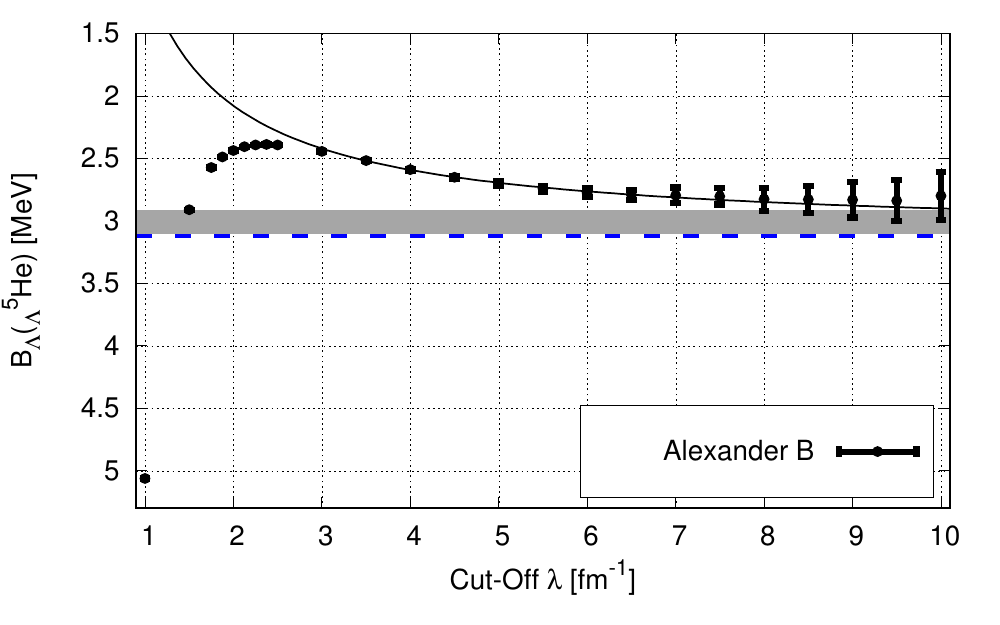}
\end{figure} 
\begin{figure}[hpt!] 
\includegraphics[width=0.5\textwidth]{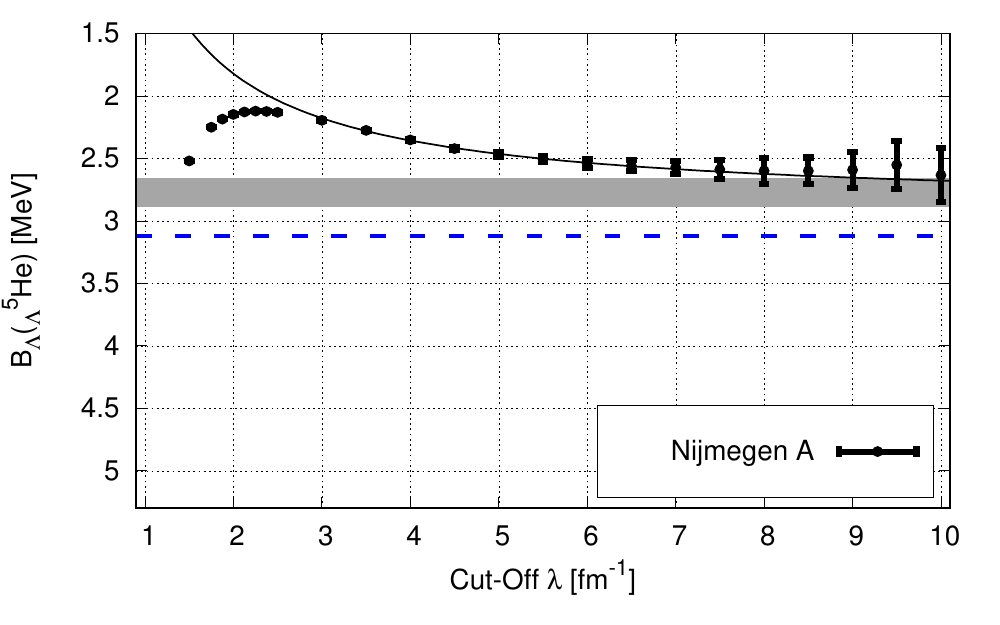} 
\includegraphics[width=0.5\textwidth]{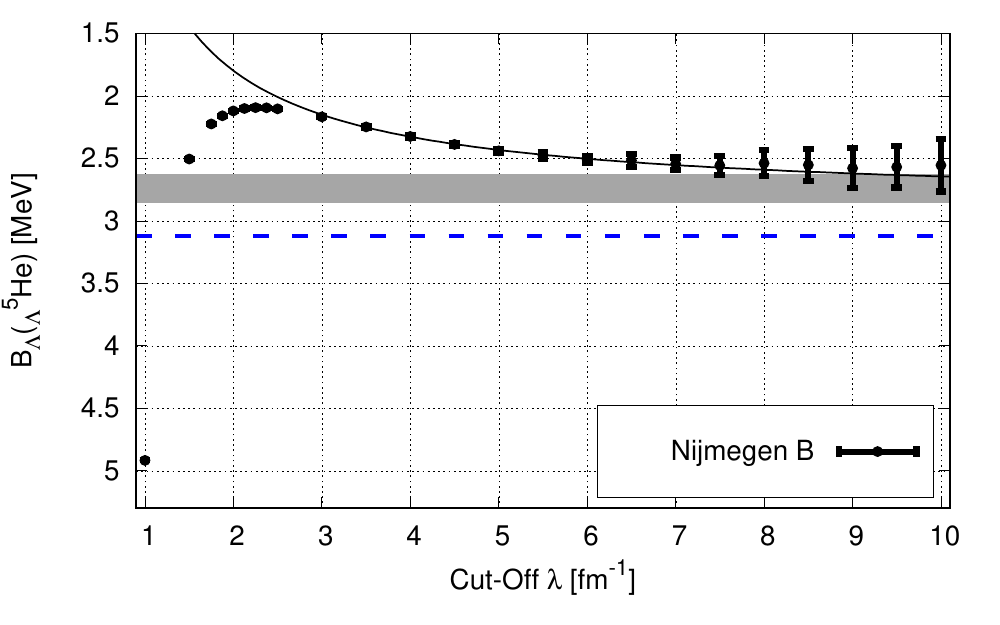}
\end{figure} 
\begin{figure}[hpt!] 
\includegraphics[width=0.5\textwidth]{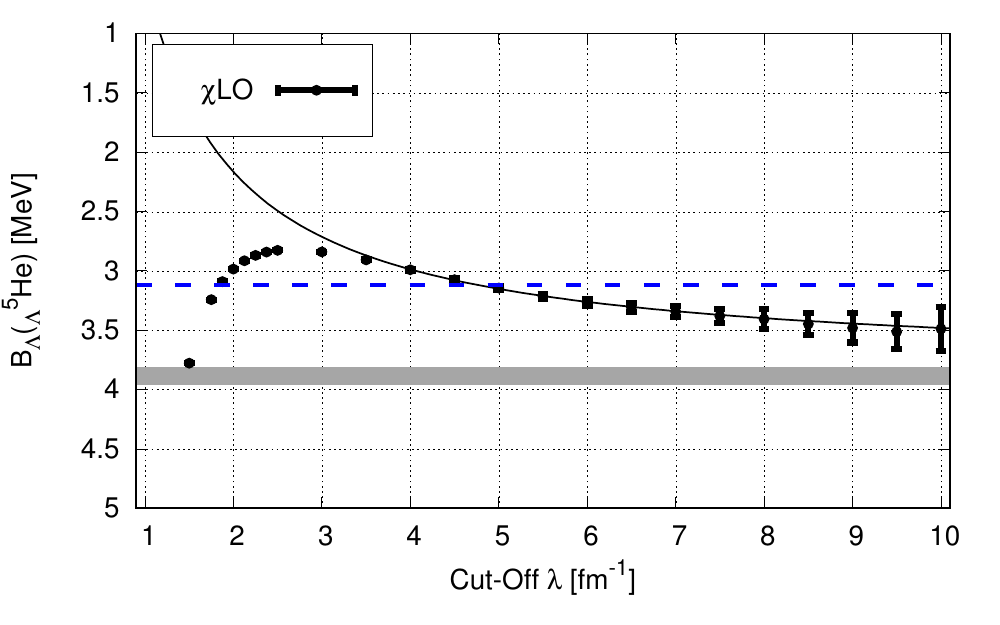} 
\includegraphics[width=0.5\textwidth]{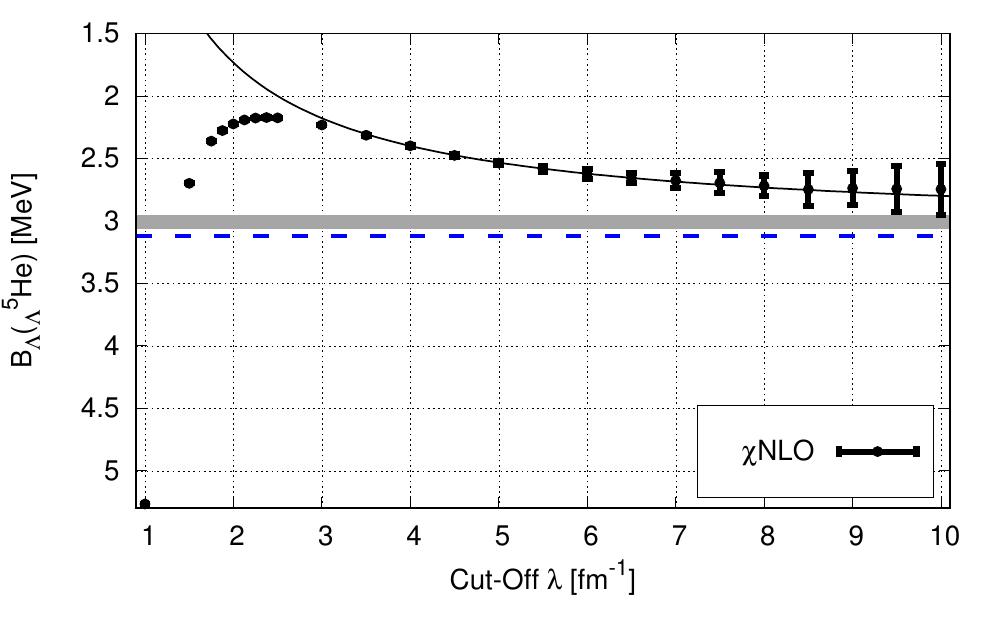}
\label{fig:Bl5} 
\caption{Dependence of $B_{\Lambda}(_\Lambda^5$He), evaluated by 
subtracting $B(^4$He) from $B(_\Lambda^5$He) in SVM calculations, on the 
\nopieft~cut-off parameter $\lambda$. The black error bars are estimated 
as the quadratic sum of of the numerical errors of the two calculations. 
The blue dashed lines represent $B_{\Lambda}^{\rm exp.}(_\Lambda^5$He), 
and the gray bands represent the two-parameter extrapolation 
$B^\lambda_\Lambda=B_\Lambda^{(\infty)}+a/\lambda$ for $\lambda\rightarrow
\infty$, fitted starting from $\lambda$=4 fm$^{-1}$. These bands include the 
numerical and extrapolation errors while the truncation error of the theory at 
LO, estimated as $\sim$300 keV, is not included. The data show modest cut-off 
dependence and RG invariance for $\lambda\geq$3~fm$^{-1}$. The extrapolated 
$B_{\Lambda}(_\Lambda^5$He) values using the Alexander or the $\chi$EFT(NLO) 
parameterizations are in good agreement with $B_{\Lambda}^{\rm exp.}
(_\Lambda^5$He), whereas $\chi$EFT(LO) reproduces it better than the 
other models do for finite cut-off values $\lambda\sim$~2-4 fm$^{-1}$.
The upper first and second groups of graphs represent the results of using the 
Alexander and the Nijmegen $\Lambda N$ scattering model inputs, respectively, 
each one with two assumed values $a_s(NN)$=$-$18.63 and $-$23.72 fm, hardly 
affecting the extrapolation $B_{\Lambda}(_\Lambda^5$He) value.}
\end{figure} 

In the present work, as in \cite{mypaper}, \nopieft~was developed and applied 
to $s$-shell hypernuclei up to A$\leq$5 using the Numerov algortihm and the 
stocastic variational method (SVM) \cite{SVa98} for a range of cut-offs 
between 1 to 10 fm$^{-1}$. The deuteron binding energy $B_{\rm d}$, the 2-body 
scattering lengths listed in Table \ref{tab:LN}, $B(^3$H), $B(^3_\Lambda$H), 
$B(^4_\Lambda$H$_{\rm g.s.}$) and $B(^4_\Lambda$H$_{\rm exc.}$) were used to 
determine the three nuclear LECs and the five hypernuclear LECs. The $\Lambda$ 
separation energy in $^5_\Lambda$He was evaluated by subtracting $B(^4$He) 
from $B(^5_\Lambda$He) for all the cut-offs $\lambda$ considered here. 
These binding energies were found to depend only moderately on $\lambda$, 
for $\lambda\gtrsim 2$~fm$^{-1}$, exhibiting renormalization scale invariance 
in the limit $\lambda\to\infty$. For example, using $a_S(NN)=-18.63$~fm, 
one obtains in this limit $B(^4{\rm He})\to 29.2\pm 0.5$~MeV which compares 
well with $B_{\rm exp}(^4{\rm He})=28.3$~MeV, given that our \nopieft~is 
truncated at LO and that the Couolomb force should reduce $B(^4$He) further 
by about 1~MeV. Results for $B_{\Lambda}(_\Lambda^5$He) are shown in 
Figure~\ref{fig:Bl5}.

The calculated $\Lambda$ separation energies were extrapolated to infinite 
cut-off $\lambda\to\infty$ using inverse power expansion as suggested by 
\nopieft: $B^\lambda_\Lambda=B_\Lambda^{(\infty)}+a/\lambda+O\left(1/\lambda^2
\right)$. In Figure~\ref{fig:Bl5}, the extrapolated values $B_\Lambda^{(
\infty)}$ are represented by gray bands which account for the calculational 
uncertainty and the estimated systematic extrapolation error. Truncation 
errors, due to not including subleading orders in the theory, are listed 
in Table~\ref{tab:Bl5}. 

\begin{table}[!htb] 
\caption{$B_{\Lambda}(_\Lambda^5$He) values (MeV) in LO \nopieft~calculations, 
with cut-off parameters $\lambda$=4~fm$^{-1}$ and $\lambda\to\infty$ (see 
text) for several choices of $\Lambda N$ interaction model. The uncertainties 
for $\lambda$=4~fm$^{-1}$ are due to subtracting $B(^4$He) from 
$B(_\Lambda^5$He), whereas those for $\lambda\to\infty$ represent (i) the 
numerical error and systematic uncertainty from the extrapolation, and (ii) 
the LO truncation error, respectively. } 
\begin{tabular}{ccccccc} 
\hline 
$B_{\Lambda}(_\Lambda^5$He$)$ & Alexander[A] & Alexander[B] & Nijmegen[A] & 
Nijmegen[B] & $\chi$LO & $\chi$NLO \\ 
\hline 
$\lambda$=4~fm$^{-1}$ & 2.61(3) & 2.59(3) & 2.35(3) & 2.32(3) & 2.99(3) & 
2.40(3) \\ 
$\lambda \to \infty$  & 3.06(10)(30) & 3.01(10)(30) & 2.77(12)(30) & 
2.74(11)(30) & 3.96(08)(35) & 3.01(06)(30) \\ 
\hline
\end{tabular} 
\label{tab:Bl5} 
\end{table} 

The $B_{\Lambda}(_\Lambda^5$He) values shown in Fig.~\ref{fig:Bl5} vary from 
a moderate underbinding for $\lambda\geq 1.5$~fm$^{-1}$, with a maximum around 
$\lambda$=2-3~fm$^{-1}$, to few MeV of overbinding for smaller cut-offs, 
comparable with overbindings produced in other interaction models described 
in the Introduction. The graphs highlight a convergent pattern between 
$\lambda$=4 and 10 fm$^{-1}$, with a moderate cut-off dependence when 
$\lambda > 2$~fm$^{-1}$. The extrapolated result is in good agreement 
with the experimental value of $B_\Lambda(_\Lambda^5$He) upon using the 
experimentally-based Alexander ([A] or [B]) parametrization and also the 
$\chi$EFT(NLO) parametrization. The parametrization of $a_{\Lambda N}$ 
extracted from $\chi$EFT(LO) leads to overbinding of $\sim$1~MeV, while 
$a_{\Lambda N}$ based on the soft-core Nijmegen model ([A] or [B]) 
leads even to underbinding of a few hundreds of keV. We conclude 
that \nopieft~applications at LO prefer the smaller input values of 
$a_t(\Lambda N)$ from Table~\ref{tab:LN}. The extrapolations also show that 
changing slightly the nuclear input (from Alexander[A] to Alexander[B] 
or from Nijmegen[A] to Nijmegen[B]) hardly affects the final results, 
suggesting that the hypernuclear applications are only weakly affected 
by the nuclear interaction input used.

\subsection{Relevant cut-offs} 

The results shown in Fig.~\ref{fig:Bl5} and Table~\ref{tab:Bl5} allow 
different interpretations than the one arrived at by extrapolating to  
$\lambda\to\infty$. While a normal \nopieft~prescription is to extrapolate 
the results to large values of $\lambda$ in order to drop residual cut-off 
dependence, the cut-off may be taken to reproduce a physically reasonable 
momentum scale between $\lambda$=2 to 3 fm$^{-1}$ which represents a mass 
scale larger than the \nopieft~breaking scale $\sim 2m_\pi$ but smaller than 
vector-meson masses starting at $\approx 4$~fm$^{-1}$. Here we disregard 
possible pseudoscalar $K$-meson exchange contributions ($m_K\approx 2.5
$~fm$^{-1}$) which are known to be insignificant \cite{MGDD85}. In this case, 
$\chi$EFT(LO) gives results closer to experiment for $_\Lambda^5$He than the 
other models do.

\begin{table}[!htb] 
\caption{Effective range values $r_0$ of $s$-wave $NN$ and $\Lambda N$ 
states, and cut-off values $\lambda_0$ that reproduce these 
listed values of $r_0$.} 
\begin{tabular}{lcccc} 
\hline 
 & $S(NN)=0$ & $S(NN)=1$ & $S(\Lambda N)=0$ & $S(\Lambda N)=1$  \\ 
\hline 
$r_0$~[fm] & 2.75~\cite{HRW06} & 1.74~\cite{HRW06} & 3.0~\cite{Alex68} & 
3.1\cite{Alex68}  \\ 
$\lambda_0$~[fm$^{-1}$] & 1.11 & 1.30 & 1.47 & 1.48  \\ 
\hline
\end{tabular} 
\label{tab:r0} 
\end{table} 

\begin{figure}[thb]  
\centering 
\begin{minipage}{0.5\textwidth} 
\centering 
\includegraphics[width=1.\textwidth]{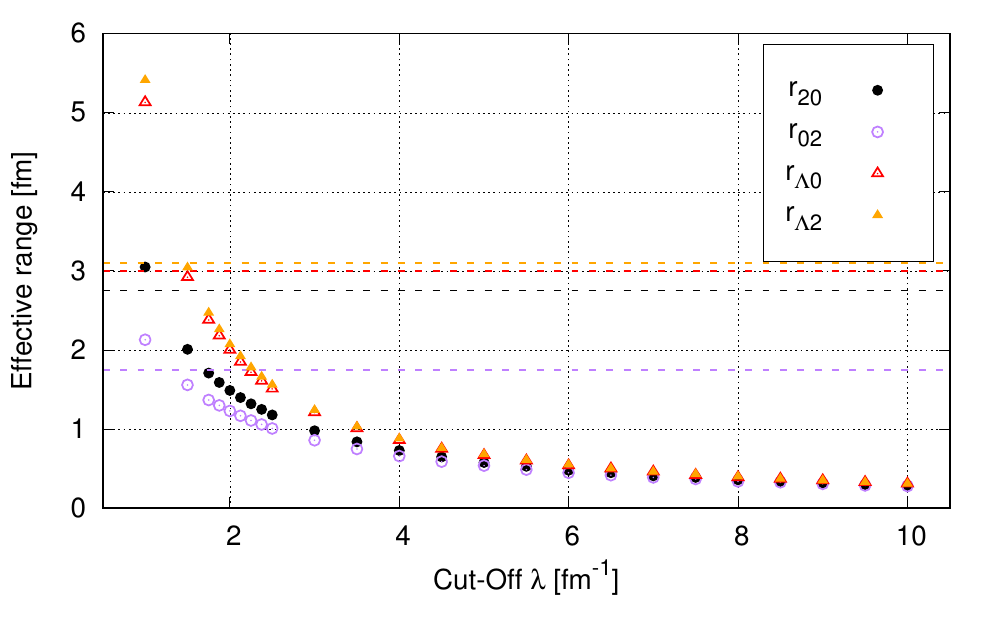} 
\end{minipage}\hfill
\begin{minipage}{0.5\textwidth} 
\caption{\textbf{a}) Effective range as a function of the cut-off parameter 
of Gaussian regulators in the four possible two-body $s$-wave systems $NN$ 
and $\Lambda N$, using the choice Alexander[B] for $a_s(NN)$. Dashed lines 
represent experimental values for the $NN$ effective range parameters 
($r_0^{20}$ in spin singlet, $r_0^{02}$ in spin triplet) and $\Lambda N$ 
effective range parameters suggested by Alexander et al. \cite{Alex68} 
($r_0^{\Lambda 0}$ in spin singlet, $r_0^{\Lambda 2}$ in spin triplet), 
all are listed in Table~\ref{tab:r0}. 
\hspace{\textwidth} 
\textbf{b}) Potential and kinetic energy expectation values as functions 
of the cut-off parameter $\lambda$ in $_\Lambda^5$He \nopieft~calculations. 
The kinetic energy and the 3-body potential energy are repulsive for all 
cut-offs, except for $\Lambda\sim 1$~fm$^{-1}$ for which the 3-body potential 
energy gives a small but attractive contribution. The kinetic (repulsive) 
energy and the 2-body potential (attractive) energy diverge as $\lambda\to
\infty$, but their sum remains finite as does the relatively small 3-body 
potential energy. }  
\centering 
\includegraphics[width=1.\textwidth]{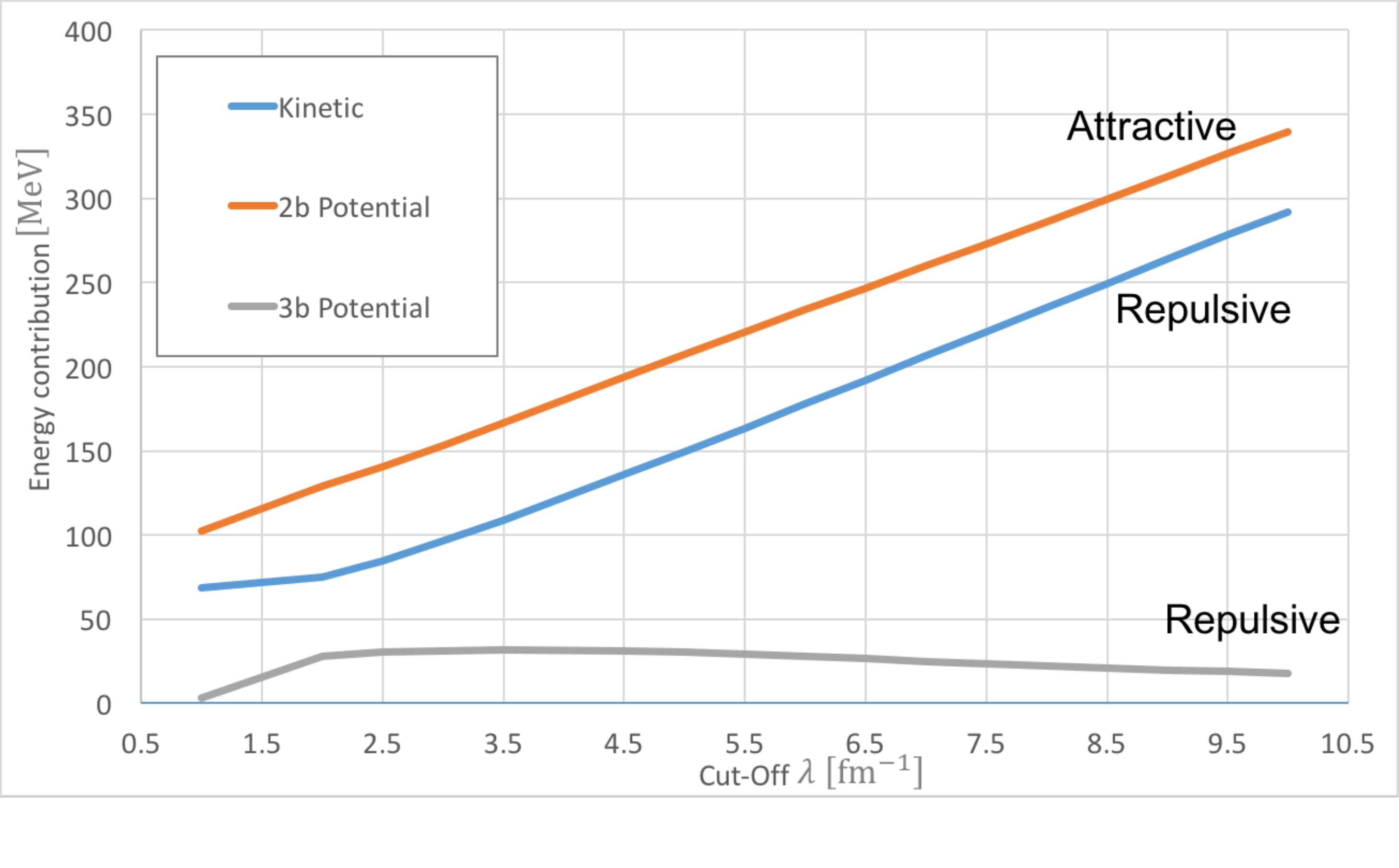} 
\end{minipage} 
\label{fig:2} 
\end{figure}

Another possibly relevant cut-off choice is one in which the regulated 
contact interaction reproduces the effective range $r_0$ of the two body 
system. According to the Wigner Bound theorem \cite{wigner,Hammer:2010fw} 
it is not possible to fix the effective range of the system with contact 
interactions in LO-\nopieft~for $\lambda\rightarrow\infty$. In fact, 
\nopieft~orders are in a one-to-one correspondence with effective-range 
expansion parameters in two body systems, the LO of the theory is associated 
with the 2-body scattering length in $s$-wave and further parameters can 
only be described by considering subleading corrections. However, this 
correspondence holds only in the limit of large cut-offs, and for each finite 
$\lambda$ it is possible to associate a finite value of the effective range. 
In Figure~\ref{fig:2}a we plot the effective ranges for the four possible 
2-body systems described in this work as a function of the cut-off $\lambda$; 
$r_0^{(20)}$ and $r_0^{(\Lambda 0)}$ represent the spin singlet effective 
range in the nuclear and hypernucler sectors, respectively, and $r_0^{(02)}$ 
and $r_0^{(\Lambda 2)}$ represent the ones in the spin triplet channel. 
The experimentally derived values of these effective ranges are listed in 
Table~\ref{tab:r0} together with cut-off values $\lambda_0$ that reproduce 
these actual effective ranges. Values of $\sim$3~fm for the $\Lambda N$ 
effective ranges, as listed in Table~\ref{tab:LN} and marked in 
Figure~\ref{fig:2}a, hold in most of the $\Lambda N$ interaction models 
listed in Table~\ref{tab:LN}. 

It is remarkable that the cut-off values $\lambda_0$ are close to the ones 
corresponding to the crossing of the $B_\Lambda(_\Lambda^5$He) curves as 
a function of $\lambda$ with the value $B_{\Lambda}^{\rm exp.}(_\Lambda^5$He) 
in Figure \ref{fig:Bl5}. Indeed, fitting the cut-off to the effective range 
practically brings in subleading-order contributions that are likely to 
enhance the predictive precision of the theory. However, this procedure is not 
guaranteed to have the same outcome for all systems, and the predictability 
regarding the truncation uncertainty of the theory is partly lost because of 
powers of $(Q/\lambda)$ which do not disappear in the expansion parameters of 
successive orders of the EFT. 

Another interesting case of cut-off is $\lambda\sim$1 fm$^{-1}$, where the 
interaction is relatively of long range, similar to that of one pion exchange. 
In this case, we recover many features arising in other calculations, such as 
a slightly attractive 3-body potential along with $^5_\Lambda$He overbound by 
as much as $1-2$~MeV.

\subsection{Energy expectation values} 

The 3-body interaction plays a fundamental role in the \nopieft~description 
of hypernuclei. On the one hand it prevents by design $A=3$ systems from 
Thomas collapse~\cite{TLH35}, and on the other hand it fine-tunes the 
interaction to reproduce the experimental data. The interaction shape becomes 
stiffer with large cut-off $\lambda$s and, as shown in Fig.~\ref{fig:2}b for 
$^5_\Lambda$He, the 2-body potential energy and kinetic energy each diverge, 
although their sum as well as the 3-body potential energy remain finite. This 
behavior might appear unexpected, since in the absence of 3-body forces the 
2-body potential energy diverges without getting fully compensated by the 
kinetic energy, thereby leading to collapse. The resolution of this paradox 
is that the 3-body interaction is extremely stiff and the sizable repulsion 
induced when three baryons move closer creates a strong correlation between 
triplets that suppresses the wavefunction, consequently reducing the 3-body 
potential energy expectation value.

\section{CONCLUSION}

In this report we reviewed and expanded on our recent application of 
\nopieft~to single $\Lambda$ hypernuclei~\cite{mypaper}. The developed theory 
extends the standard formulation of \nopieft~at LO for nuclei that uses three 
nuclear LECs, fitted to nuclear 2- and 3-body observables, by adding five new 
LECs fitted to hypernuclear 2-, 3- and 4-body observables. The two 2-body 
$\Lambda N$ LECs were determined using a wide range of input models and 
experimental data to account for the large uncertainty involved in extracting 
reliable values of the $\Lambda N$ scattering lengths. $A=3$ hypernuclear 
systems other than the observed $^3_\Lambda$H g.s. with spin $1/2$ were not 
involved in the fitting procedure while also not showing any sign of being 
bound or almost bound. In this scheme $B(^4$He) and $B(^5_\Lambda$He) are 
a prediction of the theory and are compared with experimental data. 

No alarming divergences are found in $B_\Lambda$($^5_\Lambda$He) and the 
results for $\lambda\rightarrow\infty$ differ from the experimental value 
from an underbinding of few hundred of keV to overbinding of almost 1 MeV, 
depending on the input 2-body model on which the theory is based. The theory 
shows good agreement with the experimental data for the extrapolated result 
both using the experimental parametrization and the scattering parameters 
extracted from the $\chi$EFT(NLO) model. The $\chi$EFT(LO) parametrization 
leads to a slight overbinding in the extrapolated result, but for cut-off 
values around the breaking scale ($\lambda\sim2-4$~fm$^{-1}$) it reproduces 
the experimental value more accurately than the other models do. 

We have demonstrated how it is possible to develop a \nopieft~which correlates 
the solution to the overbinding problem of single $\Lambda$ $s$-shell 
hypernuclei with tested $\Lambda N$ 2-body input parameters. The results 
suggest a larger dependence of $B_\Lambda$($^5_\Lambda$He) on the $\Lambda N$ 
spin-triplet interaction than on the spin-singlet, implying that fairly small 
values of triplet scattering lengths (roughly between $-$1.5 to $-$1.7 fm) 
are favored in order to overcome the overbinding problem. 

Lastly, we compared the kinetic energy and the 2- and 3-body potential energy 
expectation values, noticing a monotonic increase of the 2-body potential 
energy and the kinetic energy upon increasing the cut-off, while the 
three-body potential energy reaches a maximum value for cut-off values 
$\lambda\sim 5$~fm$^{-1}$. This behavior is intimately connected to the role 
played by the $NNN$ interaction in averting collapse for $A\geq 3$ and by the 
$\Lambda NN$ interaction in fine-tuning the calculated $\Lambda$ separation 
energy in $s$-shell hypernuclei.

\section{ACKNOWLEDGMENTS} 

The work of LC and NB was supported by the Pazy Foundation and by the 
Israel Science Foundation grant 1308/16.

\bibliographystyle{unsrt}
\bibliography{Thebibliography}

\begin{thebibliography}{10}

\bibitem{Alex68}
G.~Alexander, U.~Karshon, A.~Shapira, G.~Yekutieli, R.~Engelmann, H.~Filthuth,
  and W.~Lughofer.
\newblock Study of the $\ensuremath{\Lambda}${N} system in low-energy
  $\ensuremath{\Lambda}$p elastic scattering.
\newblock {\em Phys. Rev.}, 173:1452--1460, 1968.

\bibitem{NSC97}
Th.~A. Rijken, V.~G.~J. Stoks, and Y.~Yamamoto.
\newblock Soft-core hyperon-nucleon potentials.
\newblock {\em Phys. Rev. C}, 59:21--40, 1999.

\bibitem{Polinder06}
H~Polinder, J~Haidenbauer, and Ulf-G. Mei\ss{}ner.
\newblock Hyperon - nucleon interactions - a chiral effective field theory
  approach.
\newblock {\em Nucl. Phys. A}, 779:244 -- 266, 2006.

\bibitem{Haiden13}
J.~Haidenbauer, S.~Petschauer, N.~Kaiser, U.-G. Mei\ss{}ner, A.~Nogga, and
  W.~Weise.
\newblock Hyperon - nucleon interaction at next-to-leading order in chiral
  effective field theory.
\newblock {\em Nucl. Phys. A}, 915:24 -- 58, 2013.

\bibitem{sechi68}
B.~Sechi-Zorn, B.~Kehoe, J.~Twitty, and R.~A. Burnstein.
\newblock Low-energy $\ensuremath{\Lambda}$-proton elastic scattering.
\newblock {\em Phys. Rev.}, 175:1735--1740, 1968.

\bibitem{MAMI16A}
A.~Esser et~al.
\newblock Observation of $_{\mathrm{\ensuremath{\Lambda}}}^{4}\mathrm{H}$
  hyperhydrogen by decay-pion spectroscopy in electron scattering.
\newblock {\em Phys. Rev. Lett.}, 114, 2015.

\bibitem{MAMI16B}
F.~Schulz et~al.
\newblock Ground-state binding energy of $\ensuremath{_\Lambda^4}${H} from
  high-resolution decay-pion spectroscopy.
\newblock {\em Nucl. Phys. A}, 954:149 -- 160, 2016.

\bibitem{E13}
T.~O. Yamamoto et~al.
\newblock Observation of spin-dependent charge symmetry breaking in
  $\mathrm{\ensuremath{\Lambda}}${N} interaction: Gamma-ray spectroscopy of
  $_{\mathrm{\ensuremath{\Lambda}}}^{4}\mathrm{He}$.
\newblock {\em Phys. Rev. Lett.}, 115:222501, 2015.

\bibitem{Davis05}
D.H. Davis.
\newblock 50 years of hypernucl. phys.: I. the early experiments.
\newblock {\em Nucl. Phys. A}, 754:3 -- 13, 2005.

\bibitem{DHT72}
R.H. Dalitz, R.C. Herndon, and Y.C. Tang.
\newblock Phenomenological study of s-shell hypernuclei with
  $\ensuremath{\Lambda}${N} and $\ensuremath{\Lambda}${NN} potentials.
\newblock {\em Nucl. Phys. B}, 47:109 -- 137, 1972.

\bibitem{LPG14}
D.~Lonardoni, F.~Pederiva, and S.~Gandolfi.
\newblock Accurate determination of the interaction between
  $\ensuremath{\Lambda}$ hyperons and nucleons from auxiliary field diffusion
  monte carlo calculations.
\newblock {\em Phys. Rev. C}, 89:014314, 2014.

\bibitem{LGP13}
D.~Lonardoni, S.~Gandolfi, and F.~Pederiva.
\newblock Effects of the two-body and three-body hyperon-nucleon interactions
  in $\ensuremath{\Lambda}$ hypernuclei.
\newblock {\em Phys. Rev. C}, 87:041303, 2013.

\bibitem{LPL18}
D.~Lonardoni and F.~Pederiva.
\newblock {Medium-mass hypernuclei and the nucleon-isospin dependence of the
  three-body hyperon-nucleon-nucleon force}.
\newblock {\em arXiv:1711.07521v3}, 2018.

\bibitem{Wirth14}
R.~Wirth, D.~Gazda, P.~Navr\'atil, A.~Calci, J.~Langhammer, and R.~Roth.
\newblock Ab initio description of $p$-shell hypernuclei.
\newblock {\em Phys. Rev. Lett.}, 113:192502, 2014.

\bibitem{GG16A}
D.~Gazda and A.~Gal.
\newblock Ab initio calculations of charge symmetry breaking in the {A}$=4$
  hypernuclei.
\newblock {\em Phys. Rev. Lett.}, 116:122501, 2016.

\bibitem{GG16B}
D.~Gazda and A.~Gal.
\newblock Charge symmetry breaking in the {A}=4 hypernuclei.
\newblock {\em Nucl. Phys. A}, 954:161 -- 175, 2016.

\bibitem{WR18}
R.~Wirth and R.~Roth.
\newblock Light neutron-rich hypernuclei from the importance-truncated no-core
  shell model.
\newblock {\em Phys. Lett. B}, 779:336 -- 341, 2018.

\bibitem{BUC84}
A.~R. Bodmer, Q.~N. Usmani, and J.~Carlson.
\newblock Binding energies of hypernuclei and three-body
  $\ensuremath{\Lambda}\mathrm{NN}$ forces.
\newblock {\em Phys. Rev. C}, 29:684--687, 1984.

\bibitem{mypaper}
L.~Contessi, N.~Barnea, and A.~Gal.
\newblock Resolving the $\mathrm{\ensuremath{\Lambda}}$ hypernuclear
  overbinding problem in pionless effective field theory.
\newblock {\em Phys. Rev. Lett.}, 121:102502, 2018.

\bibitem{BCG15}
N.~Barnea, L.~Contessi, D.~Gazit, F.~Pederiva, and U.~van Kolck.
\newblock Effective field theory for lattice nuclei.
\newblock {\em Phys. Rev. Lett.}, 114:052501, 2015.

\bibitem{Kol99}
U.~van Kolck.
\newblock Effective field theory of short-range forces.
\newblock {\em Nucl. Phys. A}, 645:273 -- 302, 1999.

\bibitem{Bazak16}
Betzalel Bazak, Moti Eliyahu, and Ubirajara van Kolck.
\newblock Effective field theory for few-boson systems.
\newblock {\em Phys. Rev. A}, 94:052502, 2016.

\bibitem{SVa98}
Y.~Suzuki and K.~Varga.
\newblock {\em Stochastic Variational Approach to Quantum-Mechanical Few-Body
  Problems}.
\newblock Lecture Notes in Physics Monographs. Springer Berlin Heidelberg,
  1998.

\bibitem{MGDD85}
D.~J. Millener, A.~Gal, C.~B. Dover, and R.~H. Dalitz.
\newblock Spin dependence of the $\ensuremath{\Lambda}$n effective interaction.
\newblock {\em Phys. Rev. C}, 31:499--509, 1985.

\bibitem{HRW06}
R.~W. Hackenburg.
\newblock {Neutron-proton effective range parameters and zero-energy shape
  dependence}.
\newblock {\em Phys. Rev.}, C73:044002, 2006.

\bibitem{wigner}
E.~P. Wigner.
\newblock Lower limit for the energy derivative of the scattering phase shift.
\newblock {\em Phys. Rev.}, 98:145--147, 1955.

\bibitem{Hammer:2010fw}
H.~W. Hammer and Dean Lee.
\newblock {Causality and the effective range expansion}.
\newblock {\em Ann. Phys.}, 325:2212--2233, 2010.

\bibitem{TLH35}
L.~H. Thomas.
\newblock The interaction between a neutron and a proton and the structure of
  {H}$^{3}$.
\newblock {\em Phys. Rev.}, 47:903--909, 1935.

\end{thebibliography}

\end{document}